\documentclass[twocolumn,aps]{revtex4-2}
\usepackage{amsmath}
\usepackage{amssymb}
\usepackage{graphicx}
\usepackage{dcolumn}
\usepackage{bm}
\usepackage{xcolor}
\usepackage{epstopdf}
\usepackage{longtable}

\begin{document}

\preprint{Phys.Rev.B }

\title{Obstacle-Induced Gurzhi Effect and Hydrodynamic Electron Flow in Two-Dimensional Systems}
\author{A. D. Levin,$^1$ G. M. Gusev,$^1$ V. A. Chitta,$^1$ Z. D. Kvon,$^{2,3}$ A. S. Jaroshevich,$^{2}$  D.  E. Utkin,$^{2}$ D. V. Dmitriev,$^{2}$  and A. K. Bakarov$^{2}$}

\affiliation{$^1$Instituto de F\'{\i}sica da Universidade de S\~ao
Paulo, 135960-170, S\~ao Paulo, SP, Brazil}
\affiliation{$^2$Institute of Semiconductor Physics, Novosibirsk
630090, Russia}
\affiliation{$^3$Novosibirsk State University, Novosibirsk 630090,
Russia}

\date{\today}
\begin{abstract}
 The viscous flow of electrons in a narrow channel requires both strong electron-electron interactions and no-slip boundary conditions. However, introducing obstacles within the liquid can significantly increase flow resistance and, as a result, amplify the effects of viscosity. Even in samples with smooth walls, the presence of an obstacle can strongly alter electron behavior, leading to pronounced hydrodynamic effects. We investigated transport in mesoscopic samples containing a disordered array of obstacles. In contrast to samples without obstacles, which do not show a decrease in resistivity with rising temperature, samples with obstacles exhibit a significant resistivity reduction as temperature increases (the Gurzhi effect). By measuring the negative magnetoresistance, we extracted shear viscosity and other parameters through comparison with theoretical predictions. Consequently, narrow-channel samples with a disordered obstacle array provide a valuable platform for studying hydrodynamic electron flow independently of boundary conditions.
\end{abstract}
\maketitle
\section{Introduction}
The hydrodynamic approach to electron behavior in two-dimensional fermionic systems offers a unique perspective that diverges from traditional kinetic theory, revealing fascinating predictions for electron transport, particularly in small-scale samples. A key insight is that, when electron-electron interactions are strong enough, the system can be described by a viscous hydrodynamic framework, allowing for new interpretations of transport phenomena. Recent breakthroughs in materials science, especially in producing exceptionally clean samples, have enabled researchers to systematically explore these hydrodynamic effects across various two-dimensional electronic systems. Hydrodynamic electron flows are anticipated in transport phenomena when the mean free path for electron-electron collisions (denoted as $l_{ee}$) is significantly shorter than the mean free path due to impurity and phonon scattering (represented as $l$).
\begin{figure}
 \includegraphics[width=8cm]{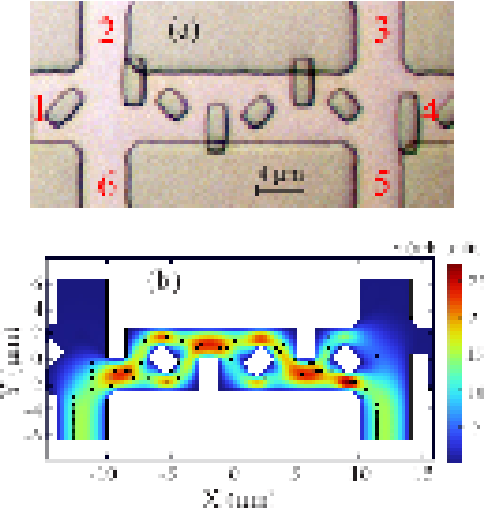}
\caption{(Color online) (a) Image of the central part of the Hall bar with 6 contacts. (b) Hydrodynamic velocity flow. Sketch of the velocity flow profile in a device with randomly oriented rectangular obstacles within a channel of narrow width of W=6 $\mu m$.  }
\end{figure}
In recent years, hydrodynamics has garnered considerable interest in the study of electronic properties within solid-state physics, leading to numerous theoretical predictions that have been experimentally confirmed \cite{narozhny, fritz}. These hydrodynamic effects include temperature-dependent resistance reduction (the Gurzhi effect) \cite{gurzhi, dejong, andreev, narozhny2, gusev1, gusev2, principi}, giant negative magnetoresistance \cite{alekseev1, narozhny3, dmitriev, raichev, gusev3, du, gusev4}, negative nonlocal resistance \cite{bandurin, torre, pellegrino2, levin}, the viscous Hall effect \cite{scaffidi, gromov, burmistrov, alekseev2, alekseev4, berdyugin, gusev5}, hydrodynamic flow around obstacles \cite{lucas, gusev6, gornyi, alekseev5, krebs}, photogenerated electron-hole plasma phenomena  \cite{pusep, pusep2} and many others.
\begin{figure*}[ht]
\includegraphics[width=18cm]{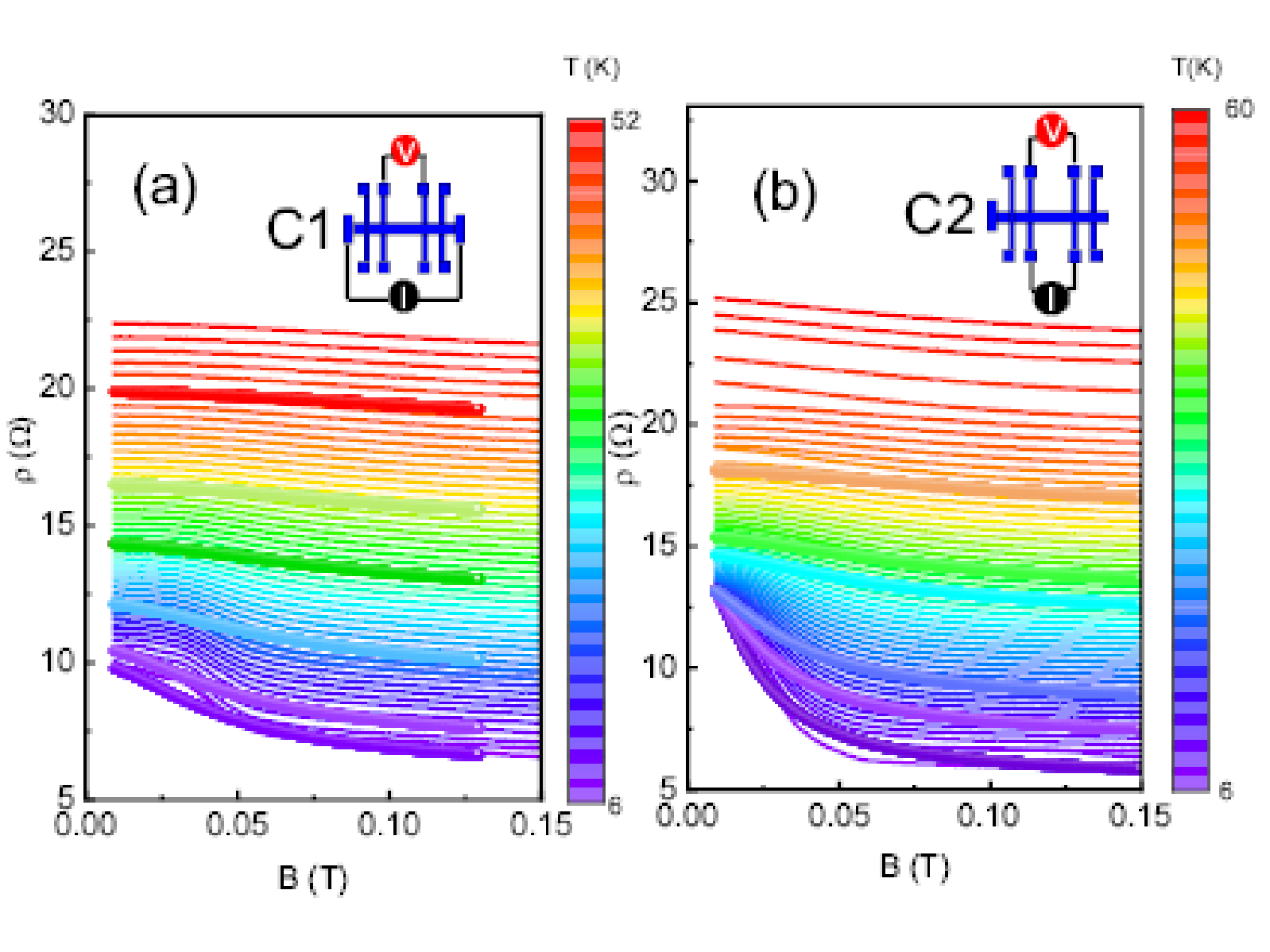}
\caption{\label{fig2}(Color online)
(a) Temperature-dependent magnetoresistivity of the sample without obstacles for configuration C1. The circles (thick lines) are examples illustrating magnetoresistance calculated from Eq. (1) for different temperatures
T (K): 6 (violet), 10 (purple),  20 (aquamarine), 31 (green),  40 (light green), 50 (red). (b) Temperature-dependent magnetoresistivity of the sample without obstacles for configuration C2. The circles (thick lines) are examples illustrating magnetoresistance calculated from Eq. (1) for different temperatures
T (K): 6 (violet),  14.9 (purple), 21 (blue violet), 36 (aquamarine), 40 (green), 50 (orange). Inserts show the configuration of the measurements. }
\end{figure*}

The reduction in resistivity with increasing temperature, initially proposed by Gurzhi, is one of the most intriguing theoretical predictions, as it appears counterintuitive at first. Normally, one might expect resistivity to rise with more frequent collisions at higher temperature. However, in narrow channels, the electron flow profile resembles Poiseuille flow, and resistivity is instead governed by the Navier-Stokes equation rather than the Boltzmann equation. As a result, it follows the relation $\rho \sim \nu \sim T^{-2}$, where $\nu =\frac{1}{4}v_{F}^2 \tau_{ee}$ is the shear viscosity ($v_{f}$ is the Fermi velocity, $\tau_{ee}$ is electron electron scattering time).

Initially, this effect was not observed in graphene or GaAs, likely due to the predominant scattering from phonons and an insufficient degree of boundary “specularity"\cite{narozhny, raichev, kiselev, raichev2}. The system’s boundary conditions can be described by diffusive scattering or slip length, denoted as $l_{s}$. In extreme cases, these conditions are classified as either “no slip” ($l_s \rightarrow 0$) or “no stress” ($l_s \rightarrow \infty$). When the slip length approaches infinity (the no-stress condition), the Gurzhi effect is not expected to appear \cite{gurzhi}. In GaAs systems, a temperature-induced resistivity decrease, attributed to the Gurzhi effect, has been observed under conditions where electrons were heated by the current \cite{dejong}, within a specialized H-shaped bar geometry \cite{gusev1} and in double \cite{gusev2} and triple quantum wells \cite{levin2}. Recent theoretical advancements have focused on adjusting slip parameters in channels with a series of narrow obstructions \cite{moessner}. This suggests that both the sample’s geometry and the specific structure of its boundaries can significantly impact transport properties, facilitating hydrodynamic behavior within confined channels \cite{keser}.

The scenario changes when impenetrable obstacles ("voids") are introduced within a narrow sample. Even "pointlike" obstacles, with radius $R < l_{ee}$, can contribute a significant hydrodynamic effect on conductivity, potentially surpassing the Drude contribution, as noted in \cite{hrushka, lucas}, regardless of the boundary conditions. For $B=0$, a circular, rigid obstacle in the hydrodynamic regime exerts a frictional force on the moving fluid, leading to the "Stokes paradox." However, this paradox is not encountered in realistic scenarios where local equilibrium is established \cite{lucas}.
The alteration of transport properties due to the presence of a circular disc obstacle has been experimentally studied in GaAs systems \cite{gusev6} and visualized in graphene samples \cite{krebs}. However, the obstacle-induced Gurzhi effect has yet to be directly observed. Thin barriers and periodic width variations in the sample also fail to enhance conditions for the Gurzhi phenomenon. Transport measurements in these structures do not exhibit the $T^{-2}$ scaling of resistivity \cite{gusev4}.

The single-obstacle scenario has been expanded to explore hydrodynamic transport in systems with random arrays of obstacles and cases featuring both rough and smooth disk edges, where electron scattering is either diffusive or specular, respectively. Within the hydrodynamic model, the resulting negative magnetoresistance—due to the suppression of dissipative viscosity—has been calculated \cite{gornyi, alekseev5}.

To amplify the effect of obstacles, we introduced an array of disordered, impenetrable rectangular obstacles within narrow channels ($6 \mu m$ width) containing two-dimensional electrons (Figure 1). We investigated  transport at $B=0$  in samples with and without these obstacles. The results revealed a notable difference: samples with obstacles exhibited a reduction in resistivity with increasing temperature (the Gurzhi effect), while samples without macroscopic scatterers showed an increase in resistance as temperature rose. Additionally, we investigated negative magnetoresistance and extracted key parameters that characterize viscous behavior and scattering, allowing us to compare these findings with theoretical predictions.
\section{Experimental results}
\begin{figure}[ht]
\includegraphics[width=8cm]{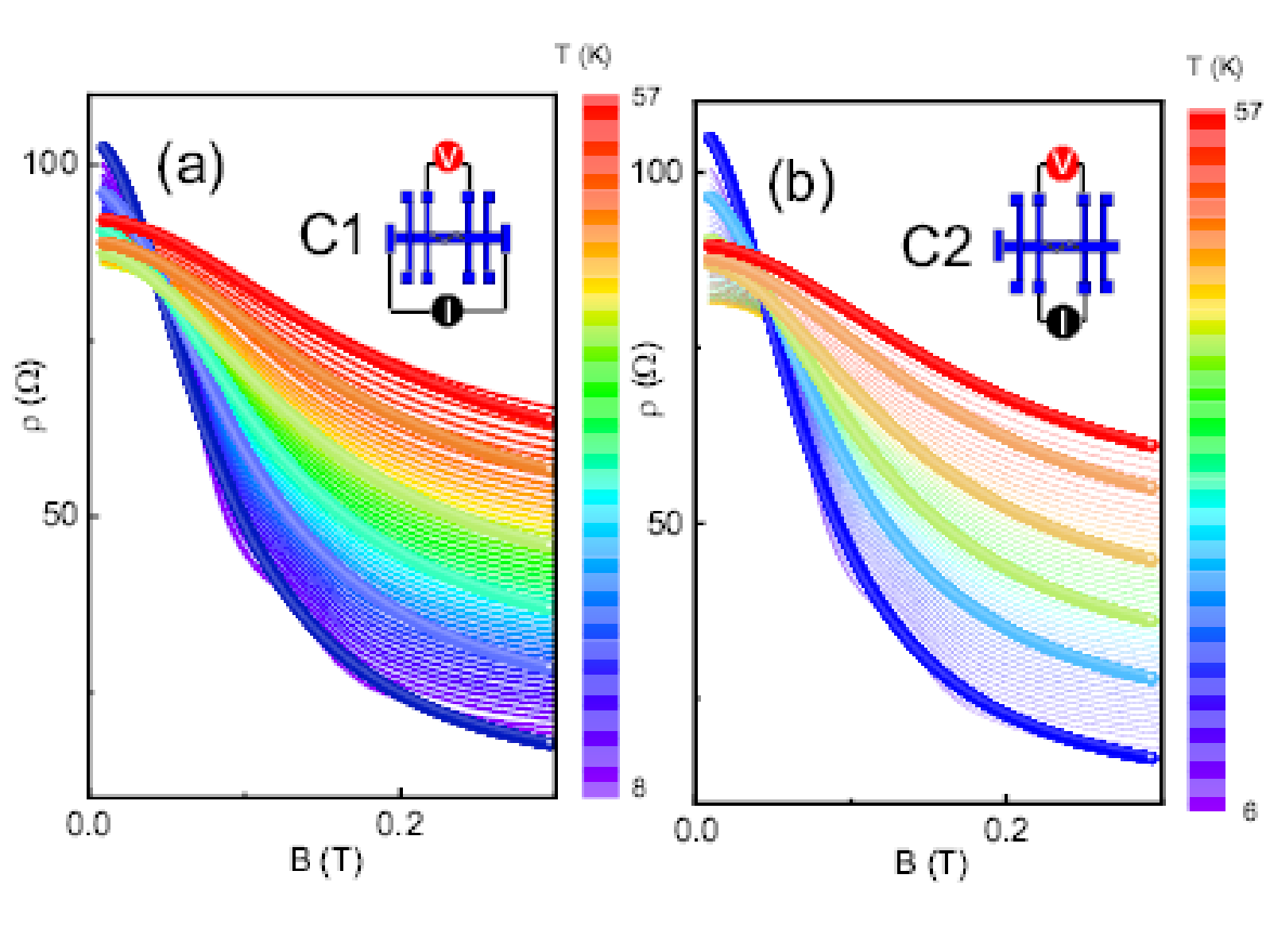}
\caption{\label{fig3}(Color online)
(a) Temperature-dependent magnetoresistivity of the sample with obstacles for configuration C1. The circles (thick lines) are examples illustrating magnetoresistance calculated from Eq. (1) for different temperatures
T (K): 6.9 (violet), 20 (blue),  31 (green),  40 (aquamarine), 50 (dark yellow), 55 (red). (b) Temperature-dependent magnetoresistivity of the sample with obstacles for configuration C2. The circles (thick lines) are examples illustrating magnetoresistance calculated from Eq. (1) for different temperatures
T (K): 6 (violet), 20 (cyan), 30 (green), 40 (yellow),  50 (dark yellow), 55(red). Inserts show the configuration of the measurements. }
\end{figure}
We fabricated our devices using high-quality GaAs quantum wells, each with a width of 14 nm and an electron density of approximately $7.1\times10^{11} cm^{-2}$ at 4.2 K. The macroscopic sample demonstrated a mobility of $2\times10^6 cm^2/(Vs)$. For our measurements, we designed a Hall bar specifically suited for multiterminal experiments, comprising of three consecutive segments with lengths of 6, 20, and $6 \mu m$, all with a width of $6 \mu m$. We also integrated eight voltage probes into this setup. Ohmic contacts to the two-dimensional electron system were created by annealing Ti/Ni/Au layers deposited on the GaAs surface.  Rectangular obstacles, sized at $2\times 1 \mu m{^2} $, were created using electron lithography followed by ion beam etching. An image of the sample with numbered contacts is displayed in Figure 1a. For comparison, we also examined unpatterned samples with identical geometry. Four samples were studied—two serving as reference patterns and two as unpatterned devices. For our measurements, we used a VTI cryostat combined with a standard lock-in detection technique to measure longitudinal resistance. To prevent overheating, we applied an alternating current (AC) in the range of $0.1-1 \mu A$, a level considered sufficiently low for these tests. The current I was directed between contacts 1 and 4, while the voltage V was measured across probes 2 and 3, yielding $R=R^{1,4}_{2,3}=V_{2,3}/I_{1,4}$(see Fig. 2a), denotes as configuration C1. In addition, we conducted measurements in an H-type configuration to enhance hydrodynamic effects \cite{gusev1}. In this case the current I was directed between contacts 6 and 5, while the voltage V was measured across probes 2 and 3, yielding $R=R^{6,5}_{2,3}=V_{2,3}/I_{6,5}$ (see Fig. 2b), denoted as configuration C2.  This study primarily focuses on magnetoresistivity measurements and zero-field resistivity behavior as a function of temperature, with a particular interest in samples with obstacles.

Figure 2 illustrates the variation in resistivity ($\rho = W/L \times R$) with respect to magnetic field strength at different temperatures for the sample without obstacles, shown under measurement configurations C1 (Figure 2a) and C2 (Figure 2b). A notable characteristic in these samples is the pronounced negative magnetoresistivity, $\rho(B) - \rho(0) < 0$, which follows a Lorentzian profile. As temperature increases, this negative magnetoresistivity diminishes in magnitude and broadens. Additionally, the resistivity at zero magnetic field rises with temperature.
\begin{figure}[ht]
\includegraphics[width=10cm]{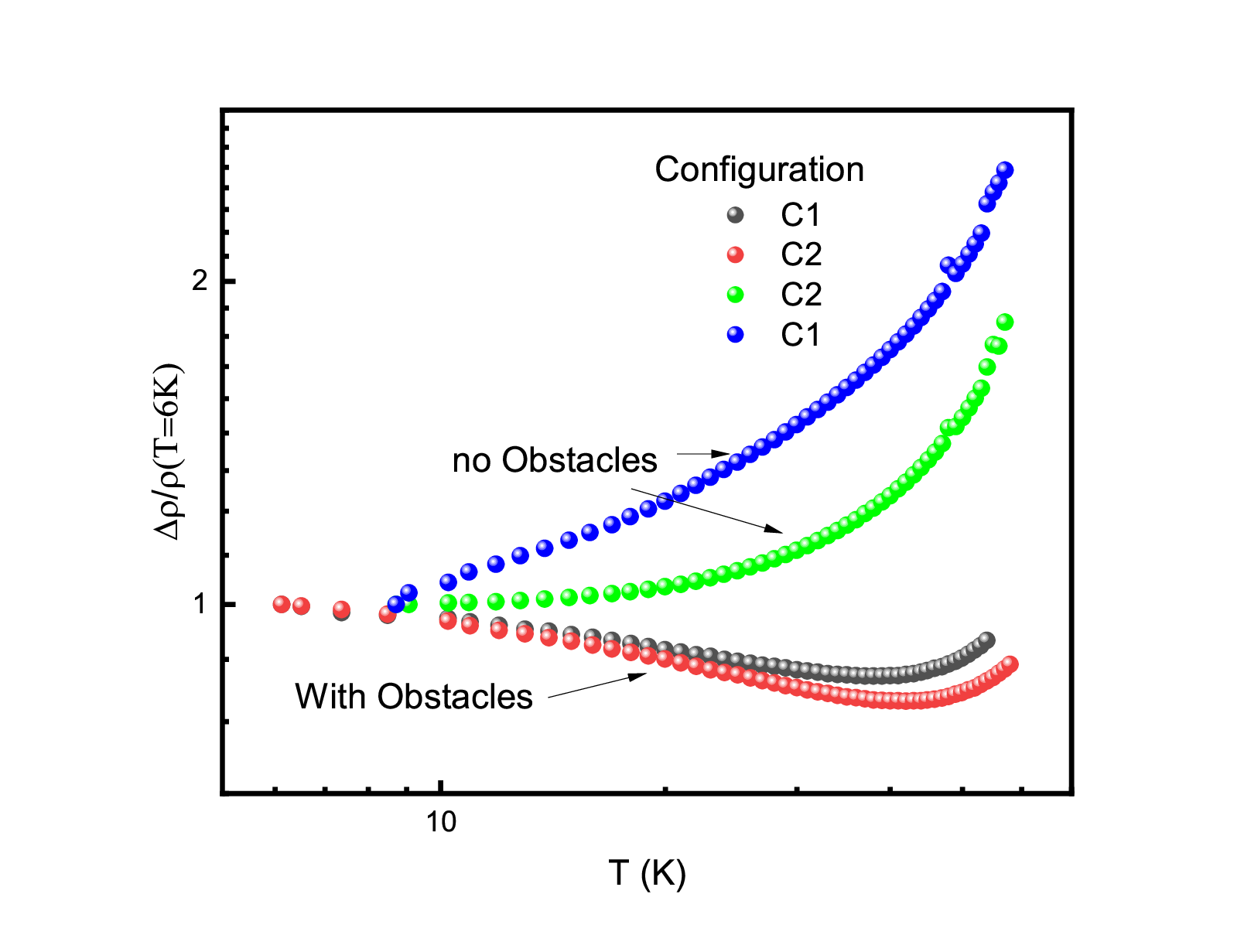}
\caption{\label{fig4}(Color online) Temperature-dependent relative resistivity of mesoscopic channels, both with and without obstacles, under different measurement configurations (C1 and C2) in the absence of a magnetic field.}
\end{figure}
In contrast, samples containing obstacles exhibit a stronger and broader Lorentzian negative magnetoresistance profile, as shown in figs 3 (a,b). More importantly, in these samples, the resistivity at zero magnetic field decreases with rising temperature for both configurations C1 and C2, which is the opposite of the behavior observed in samples without obstacles. To further highlight these differences, Figure 4 presents the temperature dependence of relative magnetoresistivity, $\Delta \rho/\rho(T=6K)$, as a function of temperature for samples with and without obstacles across both measurement configurations. First, it is noteworthy that, in samples without obstacles, a comparison of the two configurations shows a distinct behavior: in configuration C2—similar to the H-shaped configuration at temperatures below 30 K—the rate of resistivity increase is much gentler than in configuration C1. This suggests an interplay between two mechanisms: conventional phonon scattering and hydrodynamic effects. Furthermore, comparing samples with and without obstacles reveals a stark contrast in temperature response. In obstacle-free samples, resistivity more than doubles as temperature rises, while in samples with obstacles, relative resistivity significantly decreases with increasing temperature, in line with the Gurzhi effect. It is worth noting that at higher temperatures ($T > 50$ K), the resistance begins to increase with rising temperature, indicating that phonon scattering becomes a more dominant factor in resistivity compared to hydrodynamic effects above 50 K. This observation suggests that the Gurzhi effect is driven by the presence of obstacles, implying that the frictional forces around these obstacles play a substantial role in decreasing resistivity due to hydrodynamic effects in these samples. To gain a deeper understanding of this behavior, we provide a detailed comparison with theoretical models in the following section.
\section{THEORY AND DISCUSSION}
To qualitatively compare with the experimental data from samples without obstacles, we apply a model from previous research, initially designed to describe Poiseuille flow under the influence of a magnetic field \cite{alekseev1,scaffidi}.
\begin{figure}[ht]
\includegraphics[width=9cm]{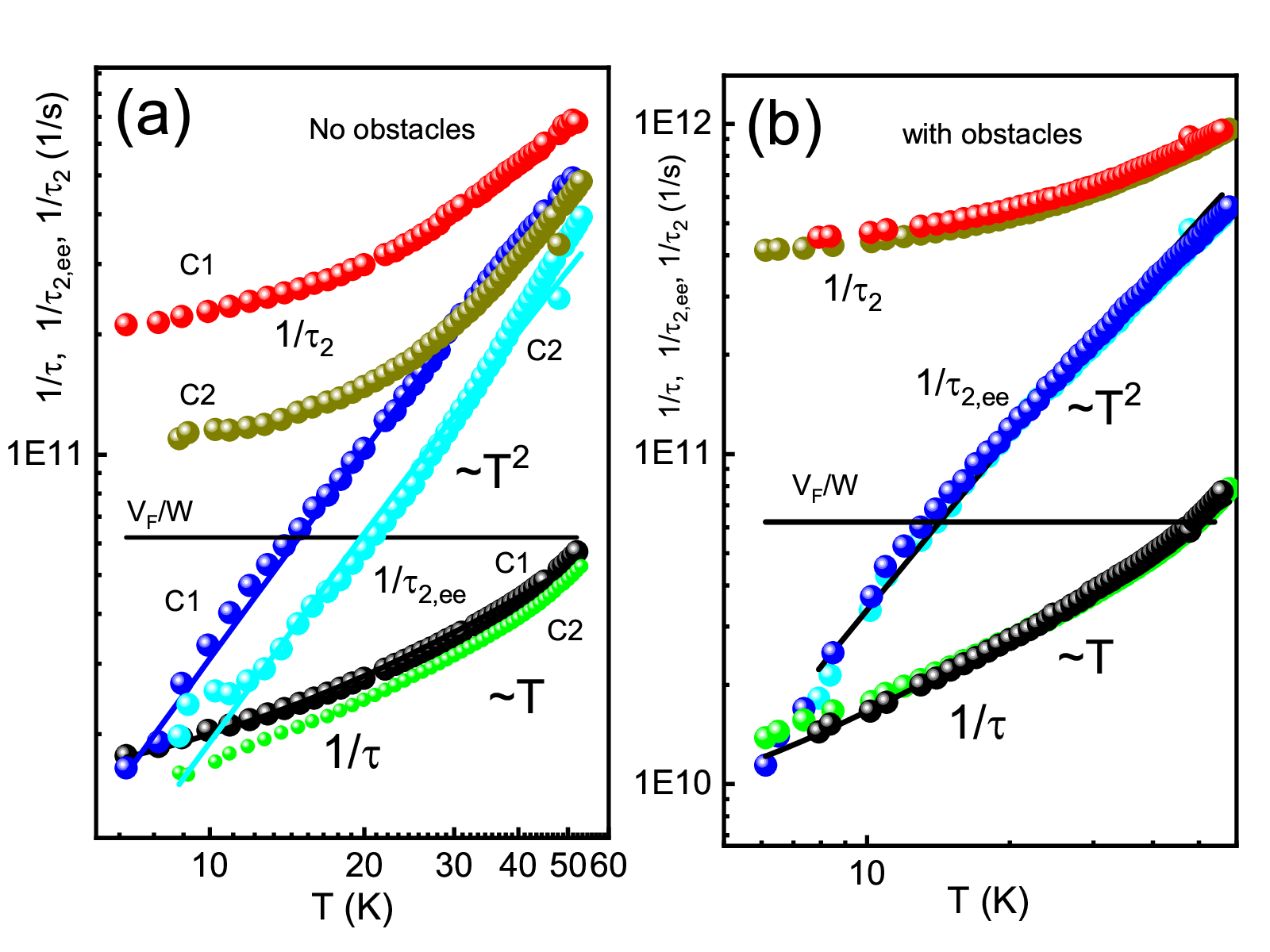}
\caption{\label{fig5}(Color online) (a) Results for sample without obstacles.  The relaxation rate,  $1/\tau_{2}$ as a function of the temperature obtained for different configurations: C1 (red), C2 (dark yellow). The relaxation rate,  $1/\tau_{2,ee}$ as a function of the temperature obtained for different configurations: C1 (blue), C2 (cyan). Solid lines- theory. The relaxation rate,  $1/\tau$ a a function of the temperature obtained for different configurations: C1 (black), C2 (green). Solid lines- theory. (b) Results for sample with obstacles. The relaxation rate,  $1/\tau_{2}$ as a function of the temperature obtained for different configurations: C1 (red), C2 (dark yellow). The relaxation rate,  $1/\tau_{2,ee}$ as a function of the temperature obtained for different configurations: C1 (blue), C2 (cyan). Solid lines- theory. The relaxation rate,  $1/\tau$ as a function of the temperature obtained for different configurations: C1 (black), C2 (green). The horizontal solid line represents the ratio $v_{F}/W$. The condition $1/\tau_{2 }> v_{F}/W >1/\tau$ corresponds to the hydrodynamic regime. }
\end{figure}
In a simplified form, the model describes resistivity as the result of two main contributions. The first stems from ballistic effects or scattering due to boundaries and defects, while the second is governed by viscosity \cite{alekseev1}.
\begin{equation}
\rho(B)= \frac{m}{e^{2}n}\left(\frac{1}{\tau}+\frac{1}{\tau^{*}}\right),\,\,\,
\end{equation}
Here, $1/\tau$ represents the scattering rate due to static disorder,  while $m=0.63m_{0}$ \cite{fu} and $n$ denote the effective mass and density, respectively, $m_0$ is the electron mass. The relaxation time $\tau^{*} = \frac{W^{*2}}{12\eta}$, where $\eta = \frac{1}{4}v_{F}^{2}\tau_{2}$ is the viscosity. The term $W^{*}$ refers to the effective sample width, which, in the case of a zero-slip boundary condition, matches the geometric width $W$. The relaxation rate $\tau_{2}$ corresponds to the shear stress relaxation time arising from electron-electron scattering. The subscript "2" signifies that the viscosity coefficient is governed by the relaxation of the second harmonic in the distribution function \cite{alekseev1}.

For a more complete formulation in magnetic field, the theory incorporates a viscosity tensor, which is dependent on the magnetic field, to determine the resistivity tensor:

\begin{equation}
\rho(B)=\left(\frac{m}{e^{2}n\tau}\right)\frac{1}{1-\tanh(\xi)/\xi}
\end{equation}
In this context, the dimensionless Gurzhi parameter is defined as $\xi = \xi_{0} \sqrt{1 + (2l_{2}/r_c)^2}$, where $\xi_{0} = W^{}/l_{G}$, with $l_{G} = \sqrt{l_{2}l}$ representing the Gurzhi length. Here, $l_{2} = v_{F} \tau_{2}$, $l = v_{F} \tau$, and $r_{c} = v_{F}/\omega_{c}$ is the cyclotron radius. The cyclotron frequency is $\omega_{c} = eB/mc$. The shear viscosity relaxation rate is given by
\begin{equation}
1/\tau_{2}(T) = 1/\tau_{2,ee}(T) + 1/\tau_{2,imp}
\end{equation}
while the momentum relaxation rate is expressed as
\begin{equation}
1/\tau(T) = 1/\tau_{0,ph}(T) + 1/\tau_{0,imp}
\end{equation}
In this expression, $1/\tau_{0,ph} = B_{ph}T$ corresponds to phonon scattering, and $1/\tau_{0,imp}$ represents scattering due to static disorder, distinct from the relaxation time for the second harmonic \cite{alekseev1}.

We then fit the magnetoresistance curves and the resistivity $\rho(T)$ at zero magnetic field,  for the unpatterned samples  and samples with obstacles  in two configurations: C1  and C2 . The fitting procedure employs three parameters: $\tau(T)$, $\tau_{2}(T)$, and the sample width $W^{*}$. Let us examine the data regarding electron-electron interactions and relaxation caused by static disorder, as derived from magnetoresistance analysis. Figure 5 illustrates the temperature-dependent behavior of the corresponding relaxation rates. To facilitate comparison with theoretical predictions, we used the parameters \( \frac{1}{\tau_{2,imp}} \), \( \frac{1}{\tau_{0,imp}} \), \( A_{ee} \), and \( B_{ph} \), as listed in Table 1. Employing Eq. (3), the e-e relaxation rate is expressed as
\begin{equation}
\frac{\hbar}{\tau_{2,ee}} = A_{ee} \frac{(kT)^2}{E_F}
\end{equation}
\begin{table}[ht]
\caption{\label{tab1} Fitting parameters of the electron system  for different configurations. Parameters are defined in the text.}
\begin{ruledtabular}
\begin{tabular}{lcccccc}
&Config. &$1/\tau_{2,imp}$&$1/\tau_{0,imp}$ & $A_{ee}$ & $B_{ph}$ &$W^{*}$   \\
& & $(10^{11} 1/s)$ & $(10^{10} 1/s)$ &   & ($10^{9} 1/(sK)$) & $\mu m$  \\
\hline
&Unpatt.,C1 & $1.95$  & $1.2$ & $1.7$ &  $0.8$ & 14\\
&Unpatt.,C2 & $0.9$  & $1.2$ & $1.05$ &  $0.8$ & 12\\
&Obstacles,C1& $4.35$  & $0.51$ & $1.8$ &  $1.15$& 2.5\\
&Obstacles, C2& $4.05$  & $0.51$ & $1.8$ &  $1.15$ & 2.5\\
\end{tabular}
\end{ruledtabular}
\end{table}
It can be observed that all relaxation rates for both samples and configurations converge onto universal curves: \( \frac{1}{\tau_{2,ee}} \sim T^2 \) and \( \frac{1}{\tau} \sim T \). However, in samples without obstacles, differences in all relaxation rates between various configurations can still be observed. This phenomenon was also noted in a previous study \cite{gusev1}, where it was attributed to the inhomogeneity of the velocity field caused by geometric effects, potentially leading to a similar outcome. In this context, \( \tau^{*} \sim \frac{d^2}{\eta} \), where \( d \) represents the characteristic period of static defects or velocity inhomogeneity. In samples with obstacles, no differences are observed between the two configurations. We attribute this to the fact that, in such samples, the contribution of obstacles to the hydrodynamic behavior outweighs the effect of conventional Poiseuille flow in a homogeneous narrow channel. Another notable difference from the table is that, in samples without obstacles, the effective width is larger than the geometric width. We attribute this discrepancy to the finite slipping length caused by specific scattering at the boundaries. A theoretical model  \cite{gromov} proposes that \( W^{*2} = W(W + 6l_s) \), which indeed predicts a larger effective width for samples with a finite slipping length. By comparing with this model, we estimate \( l_s \sim 4-5 \, \mu \mathrm{m} \) in unpatterned samples.  For diffusive boundary scattering, the velocity distribution profile in the channel is parabolic, corresponding to Poiseuille flow in a liquid. The slip length is the distance where the extrapolated velocity vanishes \cite{kiselev, raichev2}. A finite slip length modifies the velocity distribution, shaping it as a "cut parabola $(W+2l_s)$" \cite{gusev2, kiselev}.
The slip length strongly depends on the boundary conditions and geometry. For example in the previous study in the samples with shorter length ($10\mu m$) and without obstacles the Gurhi effect has been observed \cite{gusev1}. The current passing around the side probes (H-heometry) significantly disturbs the electron flux, leading to greater inhomogeneity in the velocity field compared to longer segments. The longer segments across various samples have consistently failed to exhibit the Gurzhi effect. It could indeed be interesting to study the slip length independently, as it is a parameter that warrants further investigation. However, the study of slip length is beyond the scope of our current work.
In obstacle-dominated samples, the velocity distribution profile is significantly modified by the embedded obstacles, as studied in theoretical models \cite{lucas}. Nonetheless, the profile can be approximated as parabolic in the narrow regions between obstacles, allowing us to apply the magnetic field model described in paper \cite{alekseev1}.

As shown in Table 1, the effective width of the patterned sample is $W^{*}< W$, which corresponds to the average geometric width.
 Calculations based on the diffusive model  \cite{gusev4} support this value. Notably, the slipping length in this case is found to be close to zero. This observation aligns with the idea that, in patterned samples, hydrodynamic effects are strongly enhanced by obstacles, rendering the boundary conditions less significant.
\begin{figure}[ht]
\includegraphics[width=10cm]{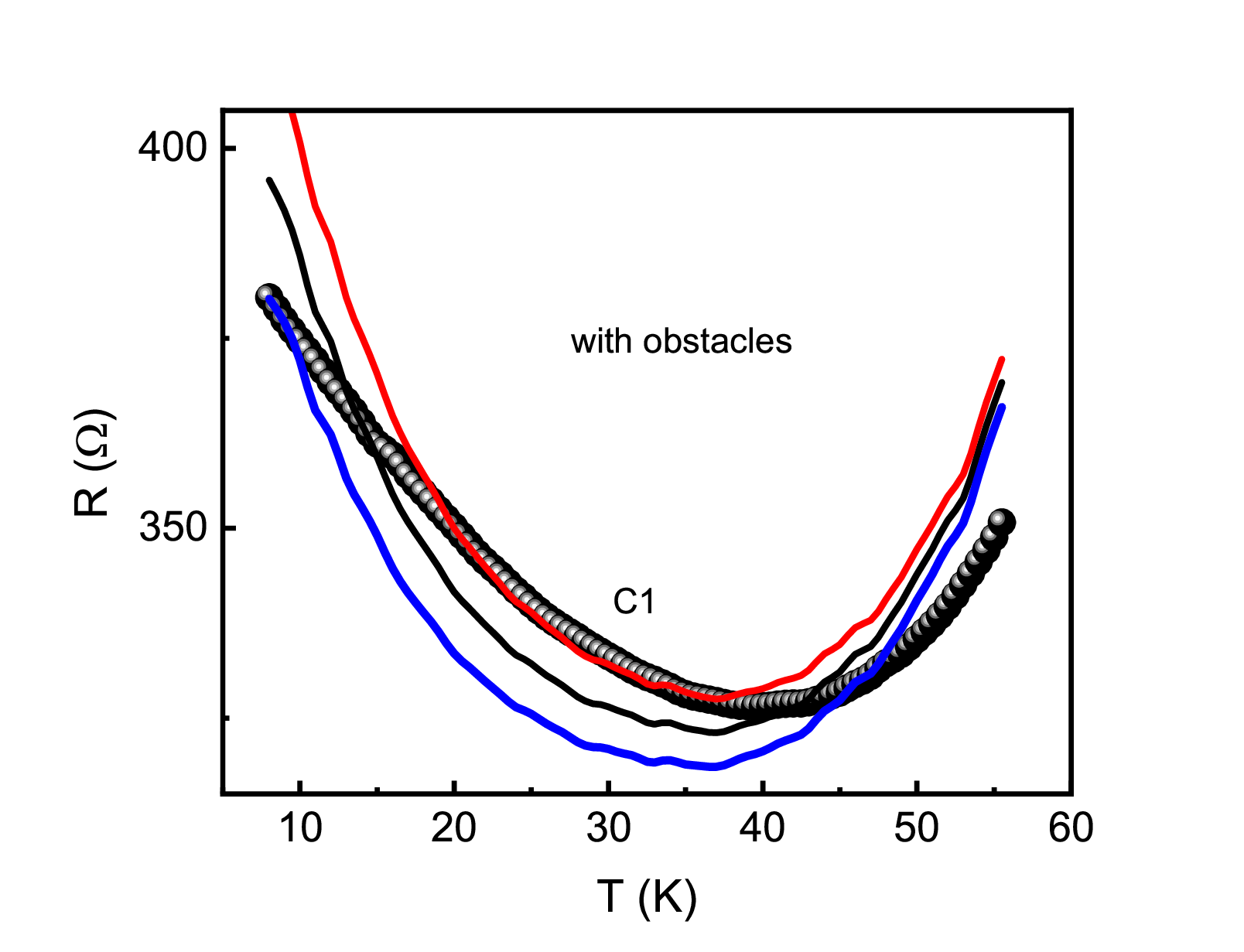}
\caption{\label{fig6}(Color online) Temperature-dependent resistance of mesoscopic channels with  obstacles for C1 configuration in the absence of a magnetic field. Solid lines represent comparison with Equations (6)-(7) for different parameters $\tau_{0, imp}$: 68 ps (black), 75 ps (red) and 61 ps (blue). }
\end{figure}
This idea is supported by a theoretical model \cite{lucas}, which examines the contribution of a single circular obstacle to resistivity. It has been demonstrated that, even in the presence of specular scattering at the boundaries, a single obstacle causes a decrease in total resistivity with increasing temperature (obstacle-induced Gurzhi effect). This effect is observed experimentally in the present study and is illustrated in Figure 4.

Based on these calculations, we can discuss the conditions for hydrodynamic effects in our samples. The hydrodynamic description is applicable under conditions where $l_{ee} < W < l$, with $l$ representing the mean free path, $W$ the width of the sample, and $l_{ee}$ the mean free path due to electron-electron collisions. However, it is important to consider the condition $l_2 < l_{ee}$, where $1/\tau_{2}=v_{F}/l_{2}$ describes the relaxation of the second harmonic of the distribution function \cite{alekseev1}. This relaxation likely involves scattering by impurities, affecting the "residual" relaxation rate of shear stress as $T \to 0$ due to electron scattering on disorder \cite{dmitriev}.

In this scenario, the condition $l_2 < W < l$ or $1/\tau_{2} > v_{F}/W > 1/\tau$ is satisfied, indicating that we remain within the hydrodynamic regime even at $T = 4.2\,\text{K}$. Experimentally, this is supported by Figure 5. Additionally, in mesoscopic samples, the scattering rate $1/\tau$ is typically lower than in macroscopic samples due to boundary scattering and geometric factors.
 It can be seen that the hydrodynamic condition $1/\tau_{2} > v_{F}/W > 1/\tau$ is satisfied across the entire temperature range used in the experiment, while the more strict condition $1/\tau_{2,ee} > v_{F}/W > 1/\tau$ is satisfied in the temperature interval $15\,\text{K} < T < 60\,\text{K}$.

It is worth noting that, despite analyzing our data using hydrodynamic theory, the interplay between ballistic and hydrodynamic effects is expected at low temperatures. The ballistic magnetoresistance in the presence of electron-electron (e-e) scattering has been considered in a theoretical model \cite{alekseev2}. For long, high-quality samples, where the skipping trajectories described above occur near the longitudinal edges, the resistance \(\rho_{xx}\) will be determined by the near-edge regions, as their contribution to the magnetic field-dependent parts of the current is greater than that of the bulk.

The relative magnetoresistance at small magnetic fields is given by the equation:
\begin{equation}
\frac{\rho(B) - \rho(0)}{\rho(0)} \sim -\frac{\omega_{c}^{2}}{\gamma^{4} W^{2} \ln\left( \frac{1}{\gamma W} \right)}
\end{equation}
where \(\gamma = 1/l_{sc}\) is the inverse scattering length due to e-e and disorder scattering, and
\[
\frac{1}{l_{\text{sc}}} = \frac{1}{l_{ee}} + \frac{1}{l}.
\]
It has been discussed in Ref. \cite{alekseev2} that for samples that are not too long, with \(W < L < 1/\gamma\), the bulk scattering rate in this formula is replaced by the reciprocal sample length \(1/L\), making the magnetoresistance temperature-independent. Moreover, it was discussed that the result \cite{alekseev2} is also applicable to  short samples with $L \sim W < l_{sc}$. In this case, the rate is replaced by \(1/W\), and the equation becomes:

\[
\frac{\rho(B) - \rho(0)}{\rho(0)} \sim - \omega_{c}^{2} W^{2}
\]

In our case, within the temperature range \( 5 \, \text{K} < T < 15 \, \text{K} \), the scattering length \( l_{sc} \) decreases from 16 \(\mu m\) to 5 \(\mu m\) with T increase. As a result, at low temperatures, we are in the short-sample regime, where ballistic magnetoresistance is expected to be temperature-independent. At higher temperatures, a transition to a ballistic + hydrodynamic regime should be observed, which then evolves into a purely hydrodynamic regime.  It would be interesting to study ballistic magnetoresistance over a broader range of sample lengths and observe the transition from temperature-dependent to temperature-independent magnetoresistance at lower temperatures, where ballistic effects dominate. This comparison could provide deeper insights into both ballistic and hydrodynamic transport. However, for the current paper, we have limited our study to a comparison with hydrodynamic theory. It is important to emphasize that, as demonstrated in the theory, the ballistic effect alone cannot replicate the Gurzhi effect and cannot explain the decrease in resistance at zero magnetic field with increasing temperature.

It is also important to note that the condition $l > W$ is quite strict and implies a pure hydrodynamic regime, leading to
\[
\rho = \frac{m}{ne^2} \eta \frac{12}{W^2}
\]
(Gurzhi regime). However, more recent theories account for an additional term in the resistance equation (1), which includes relaxation by phonons or impurities (first harmonic relaxation). Therefore, the condition $W \approx l$ is also relevant for studying viscosity.

Now, we return to the temperature dependence of resistance in the presence of obstacles, under zero magnetic field. The model proposed in \cite{lucas} predicts that obstacles enhance the total resistance as follows:
\begin{equation}
R_{\text{total}} = R_{0} + R_{\text{obst}},
\end{equation}
where \( R_{0} \) represents the resistance in the absence of obstacles, and
\[
R_{\text{obst}} = c R_{0} \frac{a_{eff}^{2}}{L^{2}},
\]
with \( c \) being a geometric factor and $a_{eff}$ effective   radius of the obstacle.

An intriguing outcome of the Stokes effect is its significant deviation from Ohmic behavior: the effective radius of the obstacle, \( a_{\text{eff}} \), is always much larger than its geometric radius $a_0$, \( a_{\text{eff}} \gg a_{0} \) \cite{lucas}. Furthermore, in the hydrodynamic regime, \( a_{\text{eff}} \) is predicted to depend only logarithmically on the actual radius, resulting in the obstacle resistance \( R_{\text{obst}} \) decreasing relatively rapidly as the temperature increases.
Let us focus on a detailed comparison between our results and this theoretical model. The theory presented in \cite{lucas} predicts a general expression for the effective obstacle radius, applicable across all transport regimes:

\begin{eqnarray}
a_{\text{eff}}^{2} \approx l_{\text{eff}} l_{2} \Bigg\{
\left(1 - \frac{2l_{\text{eff}}}{l_{2}}\right) \times \nonumber \\
\times \log \left[\frac{l_{2}}{l_{\text{eff}}}
\left(\sqrt{1 + \left(\frac{2l_{\text{eff}}}{a_{0}}\right)^{2}} - 1\right) + 1\right] \nonumber \\
+ \sqrt{1 + \left(\frac{2l_{\text{eff}}}{a_{0}}\right)^{2}} - 1
\Bigg\}^{-1}.
\end{eqnarray}
The inverse scattering length is given by
\[
\frac{1}{l_{\text{eff}}} = \frac{1}{l_{2}} + \frac{1}{l}.
\]
From this equation, it can be observed that in the intermediate temperature range, where \( l > l_{2} \) (hydrodynamic regime), one can expect $R_{\text{obst}} \sim a_{\text{eff}}^2 \sim l_{2}^2$. Consequently, the total resistance in the hydrodynamic regime is given by $ R \sim R_{0}a_{\text{eff}}^2$. The temperature dependence of $v_F/l_2=1/\tau_{2}$ is illustrated in Figure 5b. However, at higher temperatures, the resistance \( R \) may reverse its behavior and begin to increase with temperature because \( R_{0} \) grows linearly with \( T \). We compared the model's predictions with our experimental results, as shown in Figure 6. The parameters used in this fit were derived from measurements of the unpatterned sample (Figure~5a), with the exception of the relaxation rate \( \frac{1}{\tau_{o,\text{imp}}} \), which is specified in the figure captions. Additionally, we assume a geometric radius of \( a_0 = 0.5 \, \mu\text{m} \) and a geometric factor \( c = 59.8 \).  Our analysis revealed that all parameters are identical, particularly those governing the temperature dependence, such as $A_{ee}$ and $B_{ph}$. Additionally, it is evident that the theory accurately captures the temperature dependence of the resistance. However, the specific profile of this dependence exhibits slight discrepancies, likely due to the approximate nature of the model.
\section {Conclusion}
Despite recent advances in producing samples with sufficiently high mobility and strong electron-electron interactions to meet the hydrodynamic conditions for an electron fluid, fully satisfying these conditions remains challenging. This difficulty arises because scattering at the sample boundaries is often not sufficiently diffusive. Two notable examples are GaAs and graphene mesoscopic samples.

While several theoretical works have explored this parameter \cite{kiselev, raichev2}, a detailed experimental investigation is still lacking. Conducting a comprehensive study of the slip length would be highly valuable, particularly to distinguish its contribution from that of bulk hydrodynamic properties.
Some initial attempts have been made in \cite{keser}, but no thorough theoretical comparison with existing models  has been conducted. This would likely require more controlled tuning of boundary conditions. For instance, variations in plasma and chemical etching during sample preparation could be used to modify the degree of specular scattering at the boundaries, offering an experimental approach to better understand this parameter.

In this study, we fabricated GaAs narrow channels incorporating a disordered array of obstacles. We measured the temperature dependence of the resistance in these samples and compared it to unpatterned confined channels. Remarkably, we observed a stark contrast in transport behavior: samples with obstacles exhibit a significant reduction in resistivity as temperature increases (the Gurzhi effect), whereas unpatterned samples show an increase in resistance with rising temperature.

By measuring the negative magnetoresistance, we extracted shear viscosity and other parameters through comparison with theoretical predictions. Additionally, we employed a model to describe transport in samples with incorporated obstacles as a function of temperature and found reasonable agreement. These results demonstrate that narrow-channel samples with a disordered obstacle array provide a promising platform for amplifying hydrodynamic electron flow effects, independent of boundary conditions.
\section {Acknoweldgments}
We thank P.S. Alekseev for helpful discussions. This work is supported by FAPESP (São Paulo Research Foundation) Grants No. 2019/16736-2, No. 2021/12470- 8, No. 2024/06755-8, CNPq (National Council for Scientific and Technological Development), Shared Equipment Centers CKP “NANOSTRUKTURY” of the Rzhanov Institute of Semiconductor Physics SB RAS and CKP “VTAN” (ATRC) of the NSU Physics Department for the instrumental and technological support and   Russian Science Foundation (Grant No. 23-72-30003). The fabrication of structures using electron lithography was supported by the Russian Science Foundation under Grant No. 19-72-30023.


\begin{thebibliography}{41}
\bibitem{narozhny}
Boris N. Narozhny, Hydrodynamic approach to two-dimensional electron
systems, La Rivista del Nuovo Cimento, {\bf 45}, 661–736 (2022).
\bibitem{fritz}
L. Fritz and T. Scaffidi, Hydrodynamic Electronic Transport, Annu. Rev. Condens. Matter Phys.  {\bf 15} 17 (2024).
\bibitem{gurzhi}
R. N. Gurzhi, Minimum of Resistance in Impurity-free Conductors, Sov. Phys. JETP {\bf 44}, 771 (1963); Reviews of Topical Problems: Hydrodynamic Effects in Solids at Low Temperature
Sov. Phys. Usp. {\bf 11}, 255 (1968).

\bibitem{andreev}
A. V. Andreev, S. A. Kivelson, and B. Spivak, Hydrodynamic Description of Transport in Strongly Correlated Electron Systems, Phys. Rev. Lett. {\bf 106}, 256804 (2011).

\bibitem{narozhny2}
B. N. Narozhny, I. V. Gornyi, M. Titov, M. Schutt, and A. D. Mirlin, Hydrodynamics in graphene: Linear-response transport, Phys. Rev. B {\bf 91}, 035414 (2015).

\bibitem{dejong}
M. J. M. de Jong and L. W. Molenkamp, Hydrodynamic electron flow in high-mobility wires, Phys. Rev. B {\bf 51}, 13389 (1995).
\bibitem{gusev1}
G. M. Gusev, A. D. Levin, E. V. Levinson, and A. K. Bakarov, Viscous electron flow in mesoscopic two-dimensional electron gas, AIP Adv. {\bf 8},
025318 (2018).

\bibitem{gusev2}
G. M. Gusev, A. S. Jaroshevich, A. D. Levin, Z. D. Kvon  and A. K. Bakarov, Viscous magnetotransport and Gurzhi effect in bilayer electron system, Phys. Rev. B {\bf 103}, 075303 (2021).

\bibitem{principi}
A. Principi, G. Vignale, M. Carrega, and M. Polini, Bulk and shear viscosities of the two-dimensional electron liquid in a doped graphene sheet, Phys. Rev. B {\bf 93}, 125410 (2016).
\bibitem{alekseev1}
P. S. Alekseev, Negative Magnetoresistance in Viscous Flow of Two-Dimensional Electrons, Phys. Rev. Lett. {\bf 117}, 166601 (2016).
\bibitem{narozhny3}
B. N. Narozhny and M. Schutt, Magnetohydrodynamics in graphene: Shear and Hall viscosities, Phys. Rev. B {\bf 100}, 035125 (2019).
\bibitem{dmitriev}
P. S. Alekseev and A. P. Dmitriev, Viscosity of two-dimensional electrons, Phys. Rev. B {\bf 102}, 241409(R) (2020).
\bibitem{raichev}
O. E. Raichev, G. M. Gusev, A. D. Levin, and A. K. Bakarov, Manifestations of classical size effect and electronic viscosity in the magnetoresistance of narrow two-dimensional conductors: Theory and experiment, Phys. Rev. B {\bf 101}, 235314 (2020).
\bibitem{gusev3}
D. A. Khudaiberdiev,  G. M. Gusev,  E. B. Olshanetsky, Z. D. Kvon,  and N. N. Mikhailov, Magnetohydrodynamics and electron-electron interaction of massless Dirac fermions,
Phys. Rev. Research {\bf 3}, L032031 (2021).
\bibitem{du}
Xinghao Wang, Peizhe Jia, Rui-Rui Du, L. N. Pfeiffer, K. W. Baldwin, and K. W. West,
Hydrodynamic charge transport in an GaAs/AlGaAs ultrahigh-mobility two-dimensional electron gas, Phys. Rev. B {\bf 106}, L241302 (2022).
\bibitem{gusev4}
A. D. Levin, G. M. Gusev, A. S. Yaroshevich, Z. D. Kvon, and A. K. Bakarov,  Geometric engineering of viscous magnetotransport in a two-dimensional electron system, Phys. Rev. B  {\bf 108}, 115310 (2023).

\bibitem{bandurin}
D. A. Bandurin, I. Torre, R. Krishna Kumar, M. Ben Shalom,
A. Tomadin, A. Principi, G. H. Auton, E. Khestanova, K. S.
Novoselov, I. V. Grigorieva, L. A. Ponomarenko, A. K. Geim,
and M. Polini, Negative local resistance caused by viscous electron backflow in graphene, Science {\bf 351}, 1055 (2016).

\bibitem{torre}
I. Torre, A. Tomadin, A. K. Geim, and M. Polini, Nonlocal transport and the hydrodynamic shear viscosity in graphene, Phys. Rev. B {\bf 92}, 165433 (2015).

\bibitem{pellegrino2}
F. M. D. Pellegrino, I. Torre, and M. Polini, Nonlocal transport and the Hall viscosity of two-dimensional hydrodynamic electron liquids, Phys.Rev. B {\bf 96}, 195401 (2017).

\bibitem{levin}
A. D. Levin, G. M. Gusev, E. V. Levinson, Z. D. Kvon, and A. K. Bakarov, Vorticity-induced negative nonlocal resistance in a viscous two-dimensional electron system,
Phys. Rev. B {\bf 97}, 245308 (2018).
\bibitem{scaffidi}
T. Scaffidi, N. Nandi, B. Schmidt, A. P. Mackenzie, and J. E. Moore, Hydrodynamic Electron Flow and Hall Viscosity, Phys. Rev. Lett. {\bf 118}, 226601 (2017).
\bibitem{gromov}
L. V. Delacretaz and A. Gromov, Phys. Rev. Lett. {\bf 119}, 226602 (2017).

\bibitem{burmistrov}
I. S. Burmistrov, M. Goldstein, M. Kot, V. D. Kurilovich, and
P. D. Kurilovich, Dissipative and Hall Viscosity of a Disordered 2D Electron Gas, Phys. Rev. Lett. {\bf 123}, 026804 (2019).

\bibitem{alekseev2}
P. S. Alekseev and M. A. Semina, Ballistic flow of two-dimensional interacting electrons, Phys. Rev. B {\bf 98}, 165412 (2018).

\bibitem{alekseev4}
P. S. Alekseev and M. A. Semina, Hall effect in a ballistic flow of two-dimensional interacting particles, Phys. Rev. B {\bf 100}, 125419 (2019).
\bibitem{berdyugin}
A. I. Berdyugin, S. G. Xu, F. M. D. Pellegrino, R. Krishna
Kumar, A. Principi, I. Torre, M. Ben Shalom, T. Taniguchi,
K. Watanabe, I. V. Grigorieva, M. Polini, A. K. Geim, and
D. A. Bandurin, Measuring Hall viscosity of graphene{\textquoteright}s electron fluid, Science {\bf 364}, 162 (2019).
\bibitem{gusev5}
G. M. Gusev, A. D. Levin, E. V. Levinson, and A. K. Bakarov, Viscous transport and Hall viscosity in a two-dimensional electron system, Phys. Rev. B {\bf 98}, 161303(R) (2018).
\bibitem{lucas}
A. Lucas, Stokes paradox in electronic Fermi liquids. Phys. Rev. B {\bf 95}, 115425 (2017).
\bibitem{gusev6}
G. M. Gusev, A. S. Yaroshevich, A.D.Levin, Z. D. Kvon, A.K.Bakarov, Stokes flow around an obstacle in viscous two-dimensional electron liquid, Sci.Rep., {\bf 10}, 7860 (2020).
\bibitem{gornyi}
I. V. Gornyi and D. G. Polyakov, Two-dimensional electron hydrodynamics in a random array of impenetrable obstacles: Magnetoresistivity, Hall viscosity, and the Landauer dipole, Phys.Rev. B 108, 165429 (2023).
\bibitem{alekseev5}
 P. S. Alekseev and A. P. Dmitriev, Hydrodynamic magnetotransport in two-dimensional electron systems with macroscopic obstacles, Phys. Rev. B {\bf 108}, 205413 (2023).
\bibitem{krebs}
Zachary J. Krebs, Wyatt A. Behn, Songci Li, Keenan J. Smith, Kenji Watanabe, Takashi Taniguchi, Alex Levchenko, Victor W. Brar, Imaging the breaking of electrostatic dams in
graphene for ballistic and viscous fluids, Science {\bf 379}, 671 (2023).
\bibitem{pusep}
Yu. A. Pusep,  M. D. Teodoro, V. Laurindo, Jr., E. R. Cardozo de Oliveira , G. M. Gusev , and A. K. Bakarov, Diffusion of photoexcited holes in a viscous electron fluid, Phys. Rev. Lett., {\bf 128}, 136801 (2022).
\bibitem{pusep2}
M. A. T. Patricio, G. M. Jacobsen, M. D. Teodoro, G. M. Gusev,
A. K. Bakarov, and Yu. A. Pusep, Hydrodynamics of electron-hole fluid photogenerated in a mesoscopic two-dimensional channel, Phys. Rev. B {\bf 109}, L121401 (2024).
\bibitem{kiselev}
Egor I. Kiselev and Jorg Schmalian,  Boundary conditions of viscous electron flow, Phys. Rev. B {\bf 99}, 035430 (2019).
\bibitem{raichev2}
 O. E. Raichev,  Linking boundary conditions for kinetic and hydrodynamic description of fermion gas, Phys. Rev. B {\bf 105}, L041301 (2022).
 \bibitem{levin2}
A. D. Levin, G. M. Gusev, V. A. Chitta, A. S. Jaroshevich, and A. K. Bakarov, Bulk and shear viscosities in a multicomponent two-dimensional electron system, Phys. Rev. B {\bf 109}, L121401 (2024).
\bibitem{moessner}
R. Moessner, N. Morales-Durán, P. Surówka, and P. Witkowski,
Boundary-condition and geometry engineering in electronic
hydrodynamics, Phys. Rev. B {\bf 100}, 155115 (2019).
\bibitem{keser}
A. C. Keser, D. Q. Wang, O. Klochan, D. Y. H. Ho,
O. A. Tkachenko, V. A. Tkachenko, D. Culcer, S. Adam, I.
Farrer, D. A. Ritchie, O. P. Sushkov, and A. R. Hamilton,
Geometric Control of Universal Hydrodynamic Flow in a Two-
Dimensional Electron Fluid, Phys. Rev. X {\bf 11}, 031030 (2021).
\bibitem{hrushka}
 M. Hruska and B. Spivak, Conductivity of the classical
 two-dimensional electron gas, Phys. Rev. B 65, 033315
 (2002).
\bibitem{fu}
X. Fu,  Q. A. Ebner, Q. Shi,  M. A. Zudov,  Q. Qian,  J. D. Watson,  and M. J. Manfra, Microwave-induced resistance oscillations in a back-gated GaAs quantum well, Phys.Rev B {\bf 95}, 235415 (2017).
\end{thebibliography}
\end{document}